%% file: main.tex
\documentclass[a4paper,11pt]{article}

\usepackage{jheppub}

\input{definitions}

\hypersetup{colorlinks}
\glsdisablehyper

\newcommand{\ccite}[1]{reference \cite{#1}}

\preprint{IRMP-CP3-22-51}

\title{Exploiting exotic LHC datasets for long-lived new particle searches}

\author[a,b,c]{Hesham El Faham, }
\author[a]{Andrea Giammanco,}
\author[d]{and Jan Hajer}

\emailAdd{hesham.elfaham@manchester.ac.uk}
\emailAdd{andrea.giammanco@uclouvain.be}
\emailAdd{jan.hajer@tecnico.ulisboa.pt}

\affiliation[a]{Centre for Cosmology, Particle Physics and Phenomenology (CP3), Université Catholique de Louvain, B-1348 Louvain-la-Neuve, Belgium}
\affiliation[b]{Inter-University Institute for High Energies (IIHE), Vrije Universiteit Brussel, B-1050 Brussels, Belgium}
\affiliation[c]{Department of Physics and Astronomy, University of Manchester, Oxford Road, Manchester M13~9PL, United Kingdom}
\affiliation[d]{Centro de Física Teórica de Partículas (CFTP), Instituto Superior Técnico (IST), Universidade de Lisboa, 1049-001 Lisboa, Portugal}


\arxivnumber{2211.02171}

\abstract{
Motivated by the expectation that \NP may manifest itself in the form of very heavy new particles, most of the operation time of the \LHC is devoted to \pp collisions at the highest achievable energies and collision rates.
The large collision rates imply tight trigger requirements that include high thresholds on the final-state particles' transverse momenta \pT and an intrinsic background in the form of particle pileup produced by different collisions occurring during the same bunch crossing.
This strategy is potentially sub-optimal for several well-motivated \NP models where new particles are not particularly heavy and can escape the online selection criteria of the multi-purpose \LHC experiments due to their light mass and small coupling.

A solution may be offered by complementary datasets that are routinely collected by the \LHC experiments.
These include heavy ion collisions, low-pileup runs for precision physics, and the so-called \enquote{parking} and \enquote{scouting} datasets.
While some of them are motivated by other physics goals, they all have the usage of mild \pT thresholds at the trigger-level in common.
In this study, we assess the relative merits of these datasets for a representative model whose particular clean signature features long-lived resonances yielding displaced dimuon vertices.
We compare the reach across those datasets for a simple analysis, simulating \LHC data in \Run 2 and \Run 3 conditions with the \software{Delphes} simulation.
We show that the scouting and parking datasets, which afford low-\pT trigger thresholds by only using partial detector information and delaying the event reconstruction, respectively, have a reach comparable to the standard \pp dataset with conventional thresholds.
We also show that heavy ion and low-pileup datasets are far less competitive for this signature.
}

\begin{document}

\maketitle

\clearpage

\section{Introduction} \label{sec:introduction}

Since the inception of the \LHC, the experiments at its ring have collected a tremendous amount of data and utilised them to search for hints of physics \BSM.
Most of the run time at the \LHC and the data stored by its main experiments are dedicated to high-energy \pp collisions.
However, a fraction of time and storage are also devoted to special datasets motivated by specific physics goals.
Examples include heavy ion runs (nucleus-nucleus or proton-nucleus collisions), motivated by the study of the quark-gluon plasma and other high-energy nuclear physics phenomena, and low-luminosity \pp runs designed to provide clean events for precise measurements of some \SM parameters for which no large amounts of data are necessary.

Any dataset, regardless of its nature, is filtered by \emph{triggers}, \ie sets of selection criteria that are applied online, based on partial detector information in order to provide a sufficiently rapid decision, since storing all collision events is impossible for a hadron collider such as the \LHC.
Both \ATLAS and \CMS can store events on tape at a maximum rate of the order of one thousand events per second, shared across a large number of trigger paths.
Therefore, the bandwidth allocated to each trigger is limited, with a larger bandwidth allocated to the triggers that give more sensitivity to the studies assigned higher priority by the collaborations that operate the experiments.

As the \LHC luminosity increases, there are two basic strategies at the trigger-level to cope with the increased collision rate within a fixed bandwidth: tightening the selection criteria and, in particular, the \pT thresholds, or keeping the same thresholds but recording only a subset of the events that would otherwise pass the trigger selection.
The latter strategy is called \emph{prescaling}.
The former strategy is traditionally favoured in \BSM searches, based on the expectation that any new particles would be massive and therefore decay into very high \pT final-state particles or generate a large transverse momentum imbalance $E_T^\text{miss}$ when particles remain undetected.
Such expectation is supported by the most popular models of \NP, such as supersymmetry, extra dimensions, and gauge unification models.

Trigger thresholds of \LHC experiments are carefully optimised such that the primary stored \pp dataset is sensitive to as many \NP signals as possible.
However, it has been suggested that \NP could be generated in collisions at the \LHC but is then disregarded by the trigger requirements; see \eg \cite{Fischer:2021sqw,Borsato:2021aum,Alimena:2019zri} and references therein.
This is especially true for \BSM models that predict \emph{soft} decays, \ie which do not lead to events with large \pT particles or large $E_T^\text{miss}$.
Examples include \ALPs, \HNLs, or new gauge bosons, each with masses of an order of a few \unit{GeV} \cite{Blekman:2020hwr,Drewes:2018xma,Strassler:2006qa,Strassler:2006im,Drewes:2019fou}.
While the authors of \ccite{Blekman:2020hwr} argue for new ways of triggering for the standard dataset, we further argue that some existing triggers have sufficiently low thresholds in some non-standard datasets.
Thus, we explore whether enhancing the resources allocated to those \emph{exotic} datasets might be convenient in future runs and whether or not they can be promising in this respect.

This article reports a comparative study of the prospects to constrain this kind of \BSM physics using the standard \pp collisions with large pileup, a low-pileup \pp sample, \HI collisions, and \pp collisions saved with the scouting and parking approaches.
We rely on simulated data where the \software{Delphes} fast detector simulation \cite{deFavereau:2013fsa} is used to emulate the detector effects corresponding to a generic \LHC multi-purpose detector in \Run 2 and \Run 3 conditions.

\subsubparagraph

This article is organised as follows.
\Cref{sec:theory} presents a specific category of \BSM signals, predicted in so-called \emph{hidden sector} models, which are notoriously difficult to detect by the multi-purpose \LHC experiments, and elaborates on a particular signature (displaced dimuons) that is found in several of those models and is experimentally clean.
The alternative datasets collected by \LHC experiments are described in \cref{sec:datasets}.
\Cref{sec:simulation} provides details on the simulation of signal and background events under the conditions corresponding to the various datasets considered.
\Cref{sec:analysis} describes a simple data-analysis strategy for identifying the benchmark signature, whose results are presented in \cref{sec:results}.
Finally, we summarise the lessons learned in \cref{sec:summary}.
The predictions for the sensitivities achievable during \Run 3 are collected in \cref{sec:run 3 results} and the code used for the generation of signal and backgrounds is given in \cref{sec:code}.

\section{Low scale hidden sectors} \label{sec:theory}

Many models predict new feebly interacting light degrees of freedom, such as \ALPs, \HNLs, and additional gauge bosons, potentially as messengers to an extensive hidden sector \cite{Agrawal:2021dbo,Arina:2021nqi}.
Despite their small masses, they can have evaded detection until now due to their small couplings, which can induce tiny production cross sections and unusual long lifetimes.
However, they might appear in high-luminosity experiments or dedicated low-energy collider datasets.
Therefore, the exotic datasets considered here present an exciting opportunity to search for manifestations of such models.
For this study, we use a benchmark model with a scalar particle and adjust its coupling such that it is long-lived and therefore decays in a secondary vertex.
While the coupling strength in this specific simplified model is not particularly well-motivated on its own, the signature of a displaced decay with soft tracks generated by this simple model is a common prediction of more complicated models and a worthwhile target for dedicated searches.
Therefore, we treat this simplified benchmark model as a proxy for a class of \BSM models featuring such a signature.

\subsection{Benchmark model} \label{sec:model}

Additional particles with spin zero can solve significant problems in particle physics, such as the strong $CP$ problem of \QCD.
One famous example is the axion originating in the spontaneous breaking of the anomalous global $\U(1)$ Peccei–Quinn symmetry \cite{Peccei:1977hh}.
The associated \PNGB has a fixed relation between coupling strength, mass, and the symmetry-breaking scale.
However, here we consider more generic pseudoscalar particles, generally called \ALPs, which are not subject to this restriction.

\EFTs provide a model-independent approach to parameterising the effects of potential \NP in low-energy data \cite{Arina:2021nqi}.
One example is the \SMEFT \cite{Brivio:2017vri} built in terms of towers of operators constructed out of the \SM fields and respecting the \SM symmetries, ordered by their mass dimension.
It is possible to add the interactions of generic \PNGBs to the \SMEFT \cite{Arina:2021nqi}.
The resulting model has been published in the form of the \software{FeynRules} \cite{Alloul:2013bka} model file \software{ALPsEFT} \cite{Brivio:2017ije}.
The part of the \ALPs \EFT Lagrangian relevant to this study is
\begin{equation} \label{eq:Lagrangian}
\mathcal L_a =
\frac12 \pd_\mu a \pd^\mu a
- \frac12 m_a a^2
- c_{\tilde G} \frac{a}{f_a} G_{\mu\nu} \widetilde G^{\mu\nu}
- \i c_{a\phi} \frac{a}{f_a} \sum_f m_f (\widebar f_L f_R - \text{h.c.})
+ \dots \,,
\end{equation}
where $a$ is the \ALPs field with mass $m_a$ and decay constant $f_a$.
The coefficients of the coupling with the gluon field strength tensor $G_{\mu\nu}$ and its dual $\widetilde G^{\mu\nu} \equiv \flatfrac{\epsilon^{\mu\nu\rho\sigma}G_{\rho\sigma}}2$ is $c_{\tilde G}^{}$ (\code{CGtil}) while after \EWSB its coupling to fermions $f$ with mass $m_f$ is $c_{a\phi}$ (\code{CaPhi}).
In simple models, the decay constant $f_a$ is generically much larger than the \EW scale.
However, in \ALPs models, it suffices to adjust it such that this \NP scale is beyond the reach of current experiments $f_a\sim\order{\unit{TeV}}$.
The values of the couplings in the vicinity of the parameter space we are interested in is constrained by prior experiments.
For example, for \ALP masses of $m_{a} \lesssim \unit[60]{MeV}$ a limit on the \ALP-gluon coupling of order $\flatfrac{4c_{\tilde G}}{f_a} \lesssim \unit[10^{-5}]{GeV^{-1}}$ has been set at \unit[90]{\%} \CL and for masses of $\unit[1]{MeV} \lesssim m_a \lesssim \unit[3]{GeV}$ a limit on the \ALP-fermion interaction of the order of $\flatfrac{\abs{c_{a\phi}}}{f_a} < \unit[(10^{-8}\text{--}10^{-6})]{GeV^{-1}}$ has been set at \unit[90]{\%} \CL \cite{Brivio:2017ije}.

\section{Datasets} \label{sec:datasets}

In the following, we introduce the \LHC datasets compared in this study.
We present the actual conditions for those datasets during the \LHC \Run 2, which took place between 2015 and 2018, and those that can be realistically expected for the next few years during the recently started \Run 3 (2022--2025).
For the latter, it is important to stress that the resources allocated to non-standard datasets, meaning the running time, the bandwidth, and the data volume, depend on decisions based on the community consensus regarding the scientific priorities of the experiments and negotiations between several analysis teams.
Consequently, for \Run 3, any estimate of the future amount of events is essentially an educated guess.
Therefore, the predictions of this study for future datasets are presented in a form that is easily scalable to different amounts of data.

\subsection{Standard \pp dataset}

During most of the \LHC running time, the aim is to maximise the delivered instantaneous luminosity.
The proton beams cross every \unit[25]{ns}, and several \pp collisions occur during the same bunch crossing, leading to the \emph{pileup} of several uncorrelated collisions into a single recorded event.
This results in a nuisance for the measurements, as it obfuscates the interpretation of the events, particularly the kinematic reconstruction of decay chains.

During \Run 2, each of the two multi-purpose experiments \ATLAS and \CMS accumulated \invfb[140] at a \CM energy of $\sqrt s = \unit[13]{TeV}$.
While the average pileup for both experiments was about $35$ interactions per recorded event, there were large differences in the pileup profile between different data-taking periods \cite{ATLASlumi,CMSlumi}.
Typical muon selection thresholds are around \unit[24 and 17]{GeV} for single-muon and dimuon triggers, respectively, see \eg \cite{ATLAS:2022zwa,CMS:2022dbt}.
The goal for the recently-started \Run 3 is to accumulate about twice that amount of data at the slightly larger collision energy of $\sqrt s = \unit[13.6]{TeV}$.
Various accelerator parameters will be tuned towards that end.
One of the side effects will be an increase in the average pileup, which will roughly double with respect to \Run 2.
Although considerably lower than the very high pileup conditions expected during the runs of the \HLLHC \cite{CidVidal:2018eel}, \ie from \Run 4 onward, such increase of the average pileup in \Run3 may lead to a slight degradation of the performances of the particle identification and reconstruction algorithms of the experiments.
One example is that it may affect the discriminating power of lepton isolation and the identification of displaced vertices.

For simplicity, we approximated the pileup profile of \Run 2 by generating the corresponding \MC sample with a mean of $35$ pileup interactions per event at $\sqrt s = \unit[13]{TeV}$.
Similarly, to emulate \Run 3 data, we generated a \MC sample with a mean of $70$ interactions per event at $\sqrt s = \unit[13.6]{TeV}$.

\subsection{Low-pileup dataset}

For a minimal time during every run, the \LHC is operated on purpose at very low instantaneous luminosity to ensure low-pileup conditions.
Such an experimentally clean dataset can be optimally exploited for measurements that are not limited by statistical uncertainty but are critically affected by detector systematics.
An example of this measurement is the determination of the $W$ boson mass, a critical input to any \SM consistency check.
In the context of our study, which is related to \LLPs, the absence of pileup brings the specific advantage of ideal identification of the primary interaction vertex and precise reconstruction of secondary vertices.

The \Run 2 low-pileup dataset at $\sqrt s = \unit[13]{TeV}$ mostly consists of about \invfb[0.2] collected in a few days at the end of 2017, with an average pileup of two collisions per bunch crossing \cite{CMS:2021gwv}.
Given the negligible degradation of detector performances at such a low pileup, we approximate these conditions by simulating a dataset without any pileup.
Additional \invfb[0.3] of low-pileup data were collected at $\sqrt s = \unit[5.02]{TeV}$ in 2015 and 2017, primarily to serve as reference data for heavy ion studies but are not considered here.
In this study, we assume a muon \pT threshold of \unit[17 and 8]{GeV} for the single-muon and dimuon trigger, respectively, as applied by \CMS in the low-pileup runs of 2017 \cite{CMS:2021gwv}.

It is not easy to foresee the amount of low-pileup data that the multi-purpose experiments will be willing to accumulate in \Run 3.
The priority for this kind of data may have risen after the recent publication of the legacy \CDF measurement of the $W$ mass \cite{CDF:2022hxs}, in tension with previous measurements and therefore urgently demanding updates from the \LHC experiments.
In this study, we assume \invfb[0.5] in \Run 3, \ie the same amount of low-pileup data as during the entire \Run 2 but wholly at high energy, which, in this case, means $\sqrt s = \unit[13.6]{TeV}$.
This assumption is arbitrary and shall be understood as merely representing an order of magnitude estimate.

\subsection{Scouting dataset}

The concept of \enquote{scouting} was introduced by the \CMS experiment in \Run 1 \cite{Mukherjee:2019anz, Anderson:2016ron, Badaro:2020kkb, Duarte:2018bsd}.
After a pilot scouting run in 2011, it has been operated regularly since 2012 in \CMS.
During \Run 2, it has also been used by \ATLAS \cite{ATLAS:2019dpa,ATLAS:2018qto} (where it is called \enquote{trigger-level analysis}) and \LHCb \cite{Aaij:2019uij} (where it is known as \enquote{turbo stream}).

Scouting is based on assigning a fraction of the bandwidth to a stream of data with reduced event content.
The name comes from the possibility of \emph{scouting} those data very early for the presence of striking signs of \NP (\eg narrow resonances)
that do not require the full power of a holistic offline analysis to be identified and that would otherwise be filtered away by the tight standard triggers.
Scouting events are acquired in parallel with standard \pp data.
Therefore, they share the same accelerator conditions, particularly the amount of pileup.
Only a minimal amount of high-level information from the online reconstruction is stored for the selected events.
This implies that the reconstruction of the physics objects is less precise in this dataset than in the standard \pp one, as the online reconstruction algorithms are optimised more for speed than for resolution or other performance metrics.
However, this strategy permits a more significant fraction of events to be stored and analysed, allowing, in particular, looser trigger thresholds.

The \LHC experiments have used the scouting data to search for dijet \cite{CMS:2016ltu,ATLAS:2018qto} and dimuon \cite{CMS:2019buh} resonances, as well as \LLPs \cite{CMS:2021sch}.
Similarly to the scouting-based search reported in \ccite{CMS:2019buh}, this study is based on a two-muon final state signature; therefore, we assume the same \Run 2 integrated luminosity as utilised in that publication.
Although the \CMS scouting stream was taking data during the entire \Run 2, a high-rate dimuon trigger appropriate for the type of analysis studied in this paper was employed only in 2017 and 2018, with a dimuon \pT threshold of \unit[3]{GeV}, accumulating \invfb[96.6] \cite{Mukherjee:2019anz}.

For \Run 3, \CMS reverts to the original scouting design entirely based on particle-flow objects, increases the bandwidth, and makes the data format more offline-like.
In our study, instead of doubling the \Run 2 dataset as assumed for the standard \pp dataset, we assume a factor of three increase.
This accounts for the following factors:
\begin{itemize}
\item The dimuon trigger appropriate for this final state was not applied in all sub-periods of \Run 2, while it is likely to be used from start to end in \Run 3 and future data acquisitions.
\item Since scouting is now a more established concept whose usefulness is more broadly appreciated than just a few years ago, larger bandwidth is now allocated to it.
\end{itemize}
However, such an increase of usable scouting data from \Run 2 to \Run 3 is, to a large extent, an educated guess, as the bandwidth allocated to scouting is a tuneable parameter to be fixed according to community priorities.

\subsection{Parking dataset}

The \enquote{parking} concept consists in storing a fraction of the triggered data without running the prompt reconstruction algorithms.
The event reconstruction is delayed to later periods without data taking, when the experiments' computing resources are not used at full capacity.
The \CMS experiment applied this concept for the first time with about twelve billion events collected in 2018, of which roughly ten billion are estimated to contain $b$-quarks \cite{CMS-DP-2019-043}, which were processed after the end of \Run 2.
This dataset was specifically called \enquote{$b$-parking} because it was motivated by $b$-physics goals and therefore designed such that it mostly contained
events with $b$-quarks, with a set of triggers that included most notably a non-isolated single-muon path with a \pT threshold of \unit[12]{GeV}.
In principle, parking can be optimised for other physics studies and additional trigger paths with the same underlying approach of loosening the selection with respect to the standard \LHC dataset are being considered.

In this study, we work with a simplified definition of a $b$-parking dataset, which assumes that all the events are collected through a single-muon path with displacement but without requiring the muons to be isolated, meaning they do not have to be separated from other particles in the same event.

The number of events to be collected by parking during \Run 3 is potentially much larger compared to \Run 2, but difficult to foresee as it depends on future available computing resources and priorities within the \LHC experiments.
For \Run 3, we scale the number of events by a factor of four, assuming that for every year of running, it will be possible to park as many events as in the $b$-parking dataset of 2018.

Since resources are saturated at the start of a \LHC fill when the instantaneous luminosity and hence the pileup is most significant, events start to be parked only when the instantaneous luminosity falls below a specific value.
Therefore, pileup is, in principle, milder in the parking dataset than in the standard \pp one.
This effect is challenging to model reliably, so for simplicity, in this study, we conservatively assume that the average pileup is identical to the standard \pp dataset.

\subsection{Heavy ion dataset}

Besides the standard \pp collision, the \LHC experiments collect heavy ion data.
While primarily collected in \Pb\Pb\ collisions, heavy ion data can also be gathered from $p\Pb$ ones.
Moreover, some amount of \O\O\ and $p\O$ data are scheduled for \Run 3, and several other options are being discussed for the \HLLHC \cite{Citron:2018lsq, bruce20_HL_ion_report}.
Typically one month of every year is allocated to heavy ion operations, compared to six or seven months for \pp collisions in a typical \LHC year.
The primary purpose of these special runs is to accumulate a deeper understanding of the quark-gluon plasma known to have permeated the early universe.
Related topics in high-energy nuclear physics are also addressed with the same data.
Furthermore, it has been recently realised that the heavy ion dataset constitutes a suitable environment for a few \NP searches, for which analyses based on these data can outperform the \pp collisions-based ones \cite{Bruce:2018yzs, dEnterria:2022sut}.
For example, searches for magnetic monopoles \cite{He:1997pj, Gould:2019myj, Gould:2017zwi, MoEDAL:2021vix} and \ALPs \cite{Knapen:2016moh, Sirunyan:2018fhl, ATLAS:2020hii} have been carried out.
It has also been proposed to search for \LLPs in heavy ion collisions \cite{Drewes:2018xma, Drewes:2019vjy}, as well as in electron-ion ones \cite{Batell:2022ubw}.
Additionally, attention has been given to the searches for dark photons \cite{Goncalves:2020czp}, strangelets \cite{Angelis:2003zn,Adams:2005cu}, and sphalerons \cite{Ho:2020ltr}, also to studying $g-2$ of the $\tau$-lepton \cite{Beresford:2019gww, Dyndal:2020yen}.

Thanks to the relatively low event rate, dimuon events can be selected at the trigger level without any explicit \pT threshold (see \eg \cite{CMS:2018eso,CMS:2022sxl}), which roughly corresponds to an implicit threshold of \unit[3]{GeV} arising from the maximum track curvature that permits it to cross the detector.

\begin{table}
\begin{tabular}{lrrccl} \toprule
Dataset & \multicolumn{2}{c}{$\pT[\text{min}]/\unit{GeV}$} & \multicolumn{2}{c}{$\mathcal L_\text{int}/ \invfb$} & $\MC$ approximation \\ \cmidrule(lr){2-3} \cmidrule(lr){4-5}
& muon & dimuon & \Run 2 & \Run 3 \\ \midrule
Standard \pp & $24$ & $17$ & $140$ & $280$ & \multirow{3}{*}{Pileup of $35$ and $70$} \\
Scouting & -- & $3$ & $96.6$ & $289.8$ \\
Parking & $12$ & -- & $48.8$ & $195.2$ \\ \cmidrule{6-6}
Low-pileup & $17$ & $8$ & $0.2$ & $0.5$ & Zero pileup \\
Heavy ion & -- & $3$ & $1.6\times 10^{-6}$ & $9.6\times 10^{-6}$ & Only \Pb\Pb \\
\bottomrule \end{tabular}
\caption[Comparison of \pT thresholds and luminosities]{
Comparison of the \pT thresholds of the single muon and dimuon triggers simulated in this study, together with the integrated luminosities collected during the \LHC \Run 2 and the integrated luminosity expected after \Run 3 for all datasets considered in this paper.
Note that the integrated luminosity of the parking dataset is estimated from the number of events collected by \CMS in \Run 2 after the comparison with the $b$-quark cross section calculated using \software{Pythia}.
} \label{tab:datasets}
\end{table}

The main limitation of the heavy ion datasets is that their maximum instantaneous luminosity is orders of magnitude lower than that of \pp collisions.
This is due to the disruptive electromagnetic effects that depend on large powers of the atomic number $Z$ and therefore penalise heavy ion collisions much more than light ones.
This is discussed at length \eg in \cite{Bruce:2018yzs}, while the \LLP search proposed in \cite{Drewes:2018xma, Drewes:2019vjy} scans over several nuclear species to identify the most optimal one taking those effects into account.
Moreover, the maximum achievable \CM energy per nucleon is smaller than that of the standard \pp operations since the acceleration depends on $Z$ while the inertia depends on the mass number $A$.
The \CM energy per nucleon has been \unit[5.02]{TeV} for \Pb\Pb\ collisions during \Run 2.
At the time of writing, the \LHC experiments discuss the \CM energy to be used in \Run 3, particularly whether keeping the same \CM energy as during \Run 2 or taking data at higher energy.
The theoretical maximum that the \LHC can potentially reach is \unit[5.5]{TeV}.

Two advantages of heavy ion data partially compensate for these limitations.
The multiplicity of partonic interactions scales approximately as $A^2$, which in the case of $\prescript{208}{82}\Pb$ beams implies a factor of $\sim 4 \times 10^4$ enhancement,
\footnote{
The enhancement is even more spectacular for processes initiated by $\gamma\gamma$ interactions, as in that case, the cross sections scale as $Z^4$.
This effect has been exploited in the searches presented in \cite{MoEDAL:2021vix, Sirunyan:2018fhl, ATLAS:2020hii}.
}
and the triggers are much looser because of the low instantaneous luminosity.
Moreover, heavy ion collisions provide a low-pileup environment, yet the track multiplicity is much higher than in individual \pp collisions.
This difference in track multiplicity is largely compensated when \pp collisions include $\order{100}$ pileup interactions per event, as calculated in \cite{Drewes:2019vjy}.
The large track multiplicity does not dramatically affect the analysis performance for clean final states with muons.
In contrast, vertex multiplicity is essential if the signal muons are predicted to be displaced.

The multi-purpose experiments at the \LHC collected $\invfb[1.6 \times 10^{-6}]$ of \Pb\Pb\ data during \Run 2.
Plans for \LHC ion operation in \Run 3 foresee to reach $\invfb[9.6 \times 10^{-6}]$ \cite{bruce21_evian}.
For this study, we do not consider the impact of data collected during the collisions of other ions.

\subsubparagraph

For the datasets presented in this section, \cref{tab:datasets} summarises the lowest threshold for the single-muon and dimuon triggers together with the integrated luminosity collected during \Run 2 and the one assumed for \Run 3.
Furthermore, we give the main approximation we have applied while simulating these datasets.

\section{Monte Carlo simulation} \label{sec:simulation}

In \cref{sec:signal,sec:background}, we present the details of the \MC simulations of the signal and background events, respectively, for all the datasets.
Subsequently, we discuss the simulation of the detector effects in \cref{sec:detector}.
All adjustments to the cards are collected in \cref{sec:code}.

\subsection{Signal simulation} \label{sec:signal}

\begin{figure}
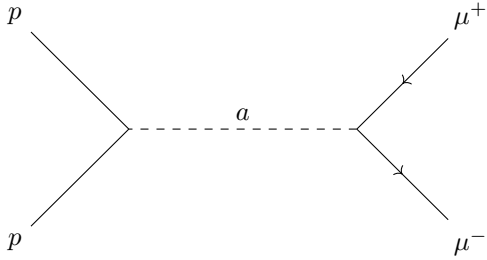

\small\includetikz*{feynman-s-channel}
\caption[Feynman diagram of the benchmark process]{
Feynman diagram of the \pp initialised \ALPs signal process introduced in \cref{sec:model}.
The \ALP constitutes a simple example for the light, feebly interacting, and long-lived new particle signatures considered in this paper.
} \label{fig:feynman diagram}
\end{figure}

The signal events are generated using the \software{FeynRules} \cite{Alloul:2013bka} model file \software{ALPsEFT} \cite{Brivio:2017ije}.
The signal process consists of an \ALP in the $s$-channel decaying to two muons as shown by the Feynman diagram in \cref{fig:feynman diagram} and is generated using \software[3.10]{MadGraph5\_aMC@NLO} \cite{Alwall:2014hca}.
The incoming $b$-quarks are chosen to be massless and included in the \PDFs through the use of the \fiveFS.
The Wilson coefficients of the operators that modify the interaction of the \ALPs to fermions and gluons are chosen to be $c_{\tilde G} = c_{a\phi} = 10^{-5}$ while the other coefficients of the model are set to zero.
The small effective coupling ensures the \ALPs are long-lived and thus displaced.
The partial width of the \ALP into its different decay channels is calculated automatically by \software{MadGraph} and we do not impose any generation-level cuts.

\paragraph{Proton collisions}

The \pp collisions are simulated at a \CM energy of $\sqrt s = \unit[13]{TeV}$ and $\sqrt s = \unit[13.6]{TeV}$ for \Run 2 and \Run 3 samples, respectively, using the \PDF \no 230000 from the \software{LHAPDF} \cite{Buckley:2014ana} set containing the NNPDF 2.3 \NLO global fit with $\alpha_s(m_Z) = 0.119$ \cite{Ball:2012cx}.

\paragraph{Heavy ion}

The heavy ion collisions are simulated at a \CM energy of $\sqrt s = \unit[5.02]{TeV}$ and $\sqrt s = \unit[5.5]{TeV}$ for \Run 2 and \Run 3 samples, respectively, using the \PDF \no 901300 from the \software{LHAPDF} set encoding the EPPS16 nuclear \PDF based on the CT14 proton \PDF at \NLO with running $\alpha_s$ for $\prescript{208}{82}\Pb$ \cite{Dulat:2015mca, Eskola:2016oht}.
Since \software{MadGraph} simulates the cross section for a single nucleon the resulting cross section has to scaled up by a factor of \eg $208^2$ in the case of lead.

\subsection{Background simulation} \label{sec:background}

Background events are simulated using \software[8.245]{Pythia} \cite{Bierlich:2022pfr} by generating $b$-jets via \QCD in $gg$ and $q\widebar q$ initial-states above a minimum \pT threshold of \unit[20]{GeV}.
This phase-space corresponds to a \pT requirement on the generated $B$-mesons.
We only consider background events that contain at least one $b$-quark and one anti-$b$-quark with final state muons.
The background generation is further optimised by imposing additional generation-level cuts:
\begin{itemize}
\item An upper limit on the vertex radius of the dimuon system of $r_v < \unit[2.5]{mm}$
\item A minimum threshold on the muons' transverse momenta of $\pT(\mu) > \unit[2]{GeV}$
\item A lower limit on the invariant mass of the dimuon system of $m_{\mu\mu} > \unit[0.5]{GeV}$
\item The muons' transverse impact parameter is required to be greater than $\abs{d_0^\mu} > \unit[0.5]{mm}$
\end{itemize}
The last cut reflects that we are primarily interested in a background environment characterised by low-\pT and high displacement and is inspired by the selection of \cite{Gershtein:2019dhy}, which was optimised for the even harsher pileup conditions expected for the \HLLHC.

\paragraph{Parking}

For the parking dataset, we generate $gg/q\widebar q \to b \widebar b$ events in \software{Pythia}.
We apply the requirement of having at least one muon or anti-muon per generated event.
The events are then passed through our analysis code which removes $\sim \unit[50]{\%}$ of the generated events.
Therefore, the cross section obtained from \software{Pythia} is scaled accordingly.
The calculated cross section can be used to translate the number of events into an equivalent integrated luminosity, and we obtain that the $10^9$ $b$-parking events stored during \Run 2 \cite{Mukherjee:2019anz} correspond to an equivalent luminosity of \invfb[48.8], see also \cref{tab:datasets}.

\paragraph{Heavy ion}

For the heavy ion background simulation, we use the Angantyr model integrated into \software{Pythia} \cite{Bierlich:2018xfw}.
Extrapolating the dynamics of \pp collisions to the ones of nuclei, this model builds up the complete hadronic final states in high energy nuclear collisions.

\subsection{Detector simulation} \label{sec:detector}

The \LHC detector effects are simulated with \software[3.4.2]{Delphes} \cite{deFavereau:2013fsa} using a modified \software{Delphes} card, based on a standard \CMS card provided with the software package.
\footnote{The \software{Delphes} cards define the parameters of the detector simulation.}
For the purpose of this study, the differences between the \ATLAS and \CMS detectors are not considered essential.
Therefore, we generalise our conclusions to both multi-purpose \LHC detectors.
\software{Delphes} allows to simulate the effect of pileup, and the user can control the emulation of necessary detector nuisances such as inefficiencies, misidentification, and loss of precision, which are adapted to the different scenarios considered in this comparative study.

\paragraph{Pileup}

The pileup events are added to the hard events during the detector simulation in \software{Delphes} using the built-in pileup card that utilises \software{Pythia}'s ability to generate soft \QCD events.
Except for some rare \SM processes, these events are designed to represent the total cross section at a hadron collider.
Out of a reservoir of $5 \times 10^4$ minimum bias events
\footnote{
Minimum bias events are defined experimentally as the most inclusive data the experiment can trigger on.
In \software{Delphes}, they are defined by a set of \software{Pythia} generator cards that comprise inelastic \pp collisions with a large cross section \cite{ATLAS:2016puo,Skands:2014pea}.
},
we add for each hard event an average of $35$ pileup events for \Run 2 simulations and $70$ pileup events for \Run 3 simulations.

\paragraph{Scouting}

As discussed in \cref{sec:datasets}, the quality of event reconstruction in the scouting dataset is reduced compared to the usual \pp runs.
Therefore, for the simulation of the scouting data, we have degraded the momentum resolution of muons according to the formula given in \cref{sec:code}, which is inspired by the degradation observed in \CMS scouting data with dimuon events \cite{CMS:2019buh}.
In that \CMS study, the degradation in $m_{\mu\mu}$ resolution depended significantly on the $\pT$ and the $\abs \eta$ of the two muons.
The \pT resolution of muons with $\pT < \unit[50]{GeV}$ is \unit[1]{\%} in the central barrel region of the detector, defined by $\abs \eta < 0.9$, and \unit[3]{\%} in the end caps of the muon system, defined by $\abs \eta > 1.2$.
A smooth interpolation between \unit[1 and 3]{\%} has been implemented in our simulation.
A cut of $\abs \eta < 1.9$ was applied in \ccite{CMS:2019buh} because higher $\abs \eta$ values lead to a much less pure sample of muon candidates, and there is not enough information at the scouting level to be able to clean the dataset.
Consequently, we apply the same cut in our analysis.
Although those details are taken from a \CMS publication, we assume that similar constraints would be motivated in any \LHC analysis.
Finally, we also impose a threshold of $\pT > \unit[3]{GeV}$ on the analysis level to account for the geometrical acceptance of the detector, which is discussed in detail in \cref{sec:analysis}.

\paragraph{Parking}

For the parking dataset, we use the track smearing module of \software{Delphes} to calculate the error on the track impact parameter.
The latter is used in defining the \IPS, which is utilised in the parking analysis, as mentioned below.

\section{Analysis} \label{sec:analysis}

\begin{figure}
\begin{panels}{2}
\includetikz{scout_dimuon_mass_run2_new_full_cuts_hybrid}
\caption{Signal} \label{fig:invariant mass signal}
\panel
\includetikz{bkg_dimuon_mass_run2_new_full_cuts_hybrid}
\caption{Background} \label{fig:invariant mass background}
\end{panels}
\caption[Dimuon invariant mass distributions]{
Probability distribution function $P_\text{norm}$ as function of the dimuon invariant mass for the signal events with nine different masses in panel \subref{fig:invariant mass signal} and the background events for the five different datasets in panel \subref{fig:invariant mass background}.
The signal distributions are shown for the scouting dataset.
The hybrid trigger thresholds are used for both the signal and background distributions.
} \label{fig:invariant mass}
\end{figure}

The baseline output of \software{Delphes} allows identifying only the primary vertex of a given event.
Since we are primarily interested in long-lived topologies and displaced vertices, we have implemented a vertexing algorithm for reconstructing the dimuon system arising from the \LLPs decay.
In this vertexing algorithm, we only take as input the tracks identified as muon candidates by the dedicated identification module in \software{Delphes}, which includes both an emulation of inefficiencies as well as fakes.
Our algorithm first identifies potential dimuon system candidates and then proceeds by imposing the \pT trigger thresholds collected in \cref{tab:datasets}.
A further refinement is attained by tightening the selection cuts used in the background simulation to $r_v < \unit[2]{mm}$ and $\abs{d_0^\mu} > \unit[1]{mm}$.
Additional requirements are needed to simulate the scouting and the parking datasets.

\paragraph{Scouting}

Given that higher values of the pseudorapidity $\abs \eta$ lead to much fewer clean muon candidates, rejecting forward candidates is necessary for a clean reconstruction of the dimuon system as mentioned in \cref{sec:detector}.
We, therefore, impose a cut of $\abs{\eta_\mu} < 1.9$ on both muon tracks as in \ccite{CMS:2019buh}.

\paragraph{Parking}

For the simulation of the parking sample, and in line with the \CMS \Lone muon trigger logic \cite{CMS-DP-2019-043}, we impose a similar requirement, namely a cut of $\abs{\eta_\mu} < 1.5$ on at least one muon candidate.
In order to further improve the trigger purity, at least one muon candidate is required to pass the lower threshold in the track \IPS, \ie $\abs{\flatfrac{d_0}{d_0^\text{err}}} > 6$ \cite{CMS-DP-2019-043}, where $d_0^\text{err}$ is the error associated to the $d_0$ measurement and calculated by \software{Delphes}.

\subsubparagraph

\begin{figure}
\begin{panels}{2}
\includetikz{scout_trackh_D0_run2_new_full_cuts_hybrid}
\caption{Signal} \label{fig:displacement signal}
\panel
\includetikz{bkg_trackh_D0_run2_new_full_cuts_hybrid}
\caption{Background} \label{fig:displacement background}
\end{panels}
\caption[Transverse impact parameter distributions]{
Probability distribution function $P_\text{norm}$ as function of the transverse impact parameter of the signal events for nine different mass points in panel \subref{fig:displacement signal} and the background events for the five different datasets in panel \subref{fig:displacement background}.
The signal distributions are shown for the scouting dataset.
The hybrid trigger thresholds are used for both the signal and background distributions.
} \label{fig:displacement}
\end{figure}

During the analysis the two muon tracks of the dimuon system are sorted according to their displacement in terms of the transverse impact parameter $\abs{d_0^\mu}$.
The dimuon candidate that has the highest displaced track is selected.
Our interest in topologies with highly displaced muon tracks motivated the analysis strategy.
\footnote{
We have checked that the number of events with more than one dimuon candidate is negligible and have therefore not specified which potential dimuon candidate must be taken in such a case.
}

\subsection{Trigger scenarios}

\begin{table}
\begin{tabular}{lccc} \toprule
 & Muon & Hybrid & Dimuon \\ \midrule
\pT threshold & \multicolumn{2}{c}{Single muon} &  Dimuon \\
\cmidrule(lr){2-3} \cmidrule(lr){4-4} \morecmidrules
\cmidrule(lr){2-2} \cmidrule(lr){3-4}
Displaced muons & $1$ & \multicolumn{2}{c}{$2$} \\
\bottomrule \end{tabular}
\caption[Trigger scenarios]{
Summary of the number of muons the \pT threshold is imposed on as well as the number of displaced muons in the three trigger scenarios used in this analysis.
The single muon and dimuon \pT thresholds for each of the datasets are given in \cref{tab:datasets}.
} \label{tab:trigger scenarios}
\end{table}

The analysis was performed for three trigger scenarios: the single muon, dimuon, and hybrid trigger scenario.
The single muon and hybrid trigger scenarios are characterised by the requirement of having a single muon candidate passing the single muon \pT threshold.
The difference between these triggers is that the single muon trigger requires only one muon to pass the displacement cut on the transverse impact parameter, \ie $\abs{d_0} > \unit[1]{mm}$.
In contrast, the hybrid trigger imposes this requirement on both muon candidates.
The dimuon trigger scenario requires both muon candidates to pass the dimuon \pT threshold and the displacement cut mentioned above.
The two trigger thresholds for the different datasets are collected in \cref{tab:datasets}, and the three trigger scenarios are summarised in \cref{tab:trigger scenarios}.

\subsubparagraph

We conclude this section by presenting the signal distributions for the mass of the dimuon system and the track impact parameter in nine different mass points in the scouting dataset using the hybrid trigger thresholds in \cref{fig:invariant mass signal,fig:displacement signal}, respectively.
These plots serve as an assessment of our vertexing algorithm's efficiency.
The background distributions for the same two observables for all the datasets and using the hybrid trigger thresholds are given in \cref{fig:invariant mass background,fig:displacement background}.
In the background distributions, aligning with our expectation, the \enquote{onia} peak is apparent at around \unit[3]{GeV}, corresponding to the $J/\psi$ meson resonances.

\section{Results and discussion} \label{sec:results}

\begin{figure}
\begin{panels}{3}
\includetikz{bpark_run2}
\caption{Parking} \label{fig:run 2 parking}
\panel
\includetikz{PU_run2}
\caption{Standard \pp} \label{fig:run 2 pileup}
\panel
\includetikz{noPU_run2}
\caption{Low-pileup} \label{fig:run 2 low-pileup}
\end{panels}
\caption[Run 2 significances: Datasets]{%
Comparison of the significance to discover the displaced low-\pT signal in arbitrary units as a function of the dimuon invariant mass.
The comparison is shown for three of the five datasets: parking \subref{fig:run 2 parking}, standard \pp \subref{fig:run 2 pileup}, and low-pileup \subref{fig:run 2 low-pileup} for the three different triggers scenarios during \Run 2.
Since only the dimuon trigger applies to the scouting and heavy ion dataset, we omit these datasets here.
} \label{fig:run 2 datasets}
\end{figure}

\begin{figure}
\begin{panels}{3}
\includetikz{single_run2}
\caption{Single muon trigger.} \label{fig:run 2 single muon}
\panel
\includetikz{hybrid_run2}
\caption{Hybrid trigger.} \label{fig:run 2 hybrid}
\panel
\includetikz{double_run2}
\caption{Dimuon trigger.} \label{fig:run 2 dimuon}
\end{panels}
\caption[\Run 2 significances: Triggers scenarios]{%
Comparison of the significance to discover the displaced low-\pT signal in arbitrary units as a function of the dimuon invariant mass.
The comparison between the relevant datasets is shown for the single muon \subref{fig:run 2 single muon}, hybrid \subref{fig:run 2 hybrid}, and dimuon \subref{fig:run 2 dimuon} trigger scenarios during \Run 2.
} \label{fig:run 2 trigger}
\end{figure}

\begin{figure}
\begin{panels}{2}
\includetikz{run2}
\caption{\Run 2} \label{fig:run 2}
\panel
\includetikz{run3}
\caption{\Run 3} \label{fig:run 3}
\end{panels}
\caption[Comparison of significances]{%
Comparison of the significance to discover the displaced low-\pT signal in arbitrary units as a function of the dimuon invariant mass.
The comparison between the relevant datasets and trigger scenarios for \Run 2 and \Run 3 are shown in panels \subref{fig:run 2} and \subref{fig:run 3}, respectively.
} \label{fig:summary}
\end{figure}

Since we do not intend to investigate a particular \NP model but rather comment on the comparative potentials to discover \NP signals across the different datasets collected at the \LHC, we refrain from giving an absolute unit for the significance.
The relevant information encoded in our results is the relative strength for discovering a displaced low-\pT signature compared between the different datasets and trigger scenarios.

The signal significance is defined as $Z = \flatfrac{s}{\sqrt{s+b}}$, where $s$ and $b$ are the numbers of signal and background events predicted in the dimuon mass window, respectively.
The width of the dimuon mass window is, in all cases, $\pm\unit[40]{\%}$ around the invariant mass $m_{\mu\mu}$ under consideration.
The predicted number of signal and background events are obtained by using scale factors $f$,
\begin{align}
s &= s_{\MC}^{} f_s, &
b &= b_{\MC}^{} f_b,
\end{align}
where $s_{\MC}^{}$ and $b_{\MC}^{}$ are the number of signal and background events counted in the corresponding \MC samples that fall within the dimuon mass window.
The $f_s$ and $f_b$ scale factors are defined by the luminosity ratios $\mathcal L_\text{data}/\mathcal L_{\MC}$.
The \MC luminosity is obtained by normalising the number of generated \MC events to the process cross section.
For example, $\mathcal L_{\MC}^\text{bkg} = \flatfrac{N_\text{gen}^\text{bkg}}{\sigma_\text{bkg}}$ where $N_\text{gen}^\text{bkg}$ was chosen in our case to be $5 \times 10^5$ of generated \MC background events and $\sigma_\text{bkg}$ is the cross section calculated by \software{Pythia}.
The luminosity of the data is the fixed integrated luminosity, $\mathcal L_\text{data}$, and is given in \cref{tab:datasets} for each type of dataset.
Although negligible, the error of the quantity $Z$ due to the size of the \MC samples is propagated and reported in the plots using,
\begin{equation}
(\delta Z)^2 = (\pd_sZ \delta s)^2 + (\pd_bZ \delta b)^2 =
\frac s4 \frac{(2 b +s)^2 f_s + b s f_b }{(b+s)^3}\,
\end{equation}
where the uncertainties on the number of signal $\delta s$ and background $\delta b$ events are taken to be the Poisson distributed errors $\sqrt{\vphantom{b}s_{\MC}^{}}$ and $\sqrt{b_{\MC}^{}}$, respectively.

\subsubparagraph

We present the comparison between the relevant trigger scenarios for three of the five datasets for \Run 2 in \cref{fig:run 2 datasets}.
We omit the scouting and heavy ion datasets since only the dimuon trigger scenario applies to these datasets.
The results for \Run 3 are given separately in \cref{fig:run 3 datasets} of \cref{sec:run 3 results}.

For $m_{\mu\mu} \geq \unit[7]{GeV}$ the single muon trigger scenario leads to the highest significance, followed by the hybrid trigger scenario.
This renders the dimuon trigger scenario the least competitive for these masses.
For signals with a large invariant mass $m_{\mu\mu}$, our observation suggests the importance of the high \pT threshold that the single muon and hybrid trigger scenarios impose on the muon candidates in comparison to the dimuon trigger scenario.
On the other hand, for $m_{\mu\mu} \leq \unit[6]{GeV}$ the hybrid trigger scenario proves the most competitive favouring the combination of the high \pT threshold along the displacement of \emph{both} muon tracks.

At this point, it is worth reminding that both the single muon and the hybrid trigger scenarios impose the same \pT threshold on the muon candidates.
Nevertheless, the former requires one displaced muon track, while the latter requires the displacement of both tracks.
This remark suggests why the hybrid trigger scenario is favoured over the single muon trigger scenario in the low-mass regions.
The resonance decay in the low-mass region is long-lived and induces a high displacement of both tracks.
In comparison, the decays in the high-mass region are prompter.
Requiring the displacement of both muon tracks in the high-mass region induces a loss of statistics arising from discarding those events in which both tracks are produced promptly.

\subsubparagraph

Having discussed the roles of the different trigger scenarios in reconstructing the dimuon signal, we now present \cref{fig:run 2 trigger} to compare the potential of each dataset in accessing such signal.
It is clear from \cref{fig:run 2 trigger} that the low-pileup and heavy ion datasets demonstrate the least competition in accessing the dimuon signal.
This is indeed expected due to the small amount of collected data compared to the other three datasets.
For the dimuon trigger scenario, the scouting dataset partially outperforms the conventional \LHC \pp dataset.
We attribute this dominance of the scouting dataset to the privileged low-\pT thresholds of the scouting triggers along the large amount of collected data.
Additionally, the parking dataset is competitive with the standard \pp dataset in the high-mass region.

We conclude this section by summarising in \cref{fig:summary} all the \LHC \Run 2 results presented in \cref{fig:run 2 datasets,fig:run 2 trigger} together with the complete overview of the \Run 3 results.
A detailed presentation of the \Run 3 results is given in \cref{sec:run 3 results}.
Here we remark that the different expected scaling for the \Run 3 luminosities enables the scouting and parking datasets to outperform the standard \pp datasets in the high-mass region.

\section{Summary and conclusion} \label{sec:summary}

\begin{figure}
\begin{panels}{3}
\includetikz{bpark_run3}
\caption{Parking} \label{fig:run 3 parking}
\panel
\includetikz{PU_run3}
\caption{Standard \pp} \label{fig:run 3 pileup}
\panel
\includetikz{noPU_run3}
\caption{Low-pileup} \label{fig:run 3 low-pileup}
\end{panels}
\caption[\Run 3 significances: Datasets]{%
Comparison of the significance to discover the displaced low-\pT signal in arbitrary units as a function of the dimuon invariant mass.
The comparison is shown for three of the five datasets: parking \subref{fig:run 3 parking}, standard \pp \subref{fig:run 3 pileup}, and low-pileup \subref{fig:run 3 low-pileup} for the three different triggers scenarios during \Run 3.
Since only the dimuon trigger applies to the scouting and heavy ion dataset, we omit these datasets here.
} \label{fig:run 3 datasets}
\end{figure}

\begin{figure}
\begin{panels}{3}
\includetikz{single_run3}
\caption{Single muon trigger.} \label{fig:run 3 single muon}
\panel
\includetikz{hybrid_run3}
\caption{Hybrid trigger.} \label{fig:run 3 hybrid}
\panel
\includetikz{double_run3}
\caption{Dimuon trigger.} \label{fig:run 3 dimuon}
\end{panels}
\caption[\Run 3 significances: Triggers scenarios]{%
Comparison of the significance to discover the displaced low-\pT signal in arbitrary units as a function of the dimuon invariant mass.
The comparison between the relevant datasets is shown for the single muon \subref{fig:run 3 single muon}, hybrid \subref{fig:run 3 hybrid}, and dimuon \subref{fig:run 3 dimuon} trigger scenarios during \Run 3.
} \label{fig:run 3 trigger}
\end{figure}

We have compared different types of datasets collected at the \LHC experiments using a simple model featuring a displaced vertex signature.
We have shown that the scouting and parking datasets can be competitive with the standard \LHC runs for such signatures in the mass region under consideration.
In contrast, the low-pileup and heavy ion datasets are limited to considerably smaller significance.

We stress that although these results have been obtained for one particular signature of a specific model, its general conclusions may be generalised to some extent.
We caution the reader that many assumptions have been made in this study and that \software{Delphes} is not meant to be an accurate detector simulation.
Therefore, some subtleties ignored in this paper may alter the relative ranking of some of the datasets considered.
However, our results indicate that scouting and parking can be very promising strategies for a large category of signals characterised by low particle masses.
Therefore, it may be worthwhile allocating significantly larger bandwidth for those datasets than in previous runs.

Moreover, our results encourage further exploitation of the unconventional datasets that have already been collected during previous \LHC runs.
We hope to see new studies that valorise these data, which may also be used as a testing ground for new ideas towards optimising the corresponding triggers for future runs.

\subsection*{Acknowledgments}

First of all, we thank Marco Drewes for contributing to the initial idea of this paper and critically accompanying its development.
Furthermore, we thank Olivier Mattelaer for the original implementation and continuous support for heavy ion collision simulation in \software{MadGraph}.
Pavel Demin and Michele Selvaggi's assistance with the \software{Delphes} software was pivotal in completing this project.
Thanks to Christian Bierlich for discussions concerning \software{Pythia}, particularly on using the Angantyr model.
We are grateful for our early talks with Simon Knapen, Steven Lowette, and Hardik Routray.
HF thanks Ken Mimasu for the technical discussions on the \software{ALPsEFT} model.
We received essential clarifications about scouting and parking in \CMS from Greg Landsberg and Maurizio Pierini.
Finally, we thank the IT team of CP3 for their assistance with the supercomputing facilities.

HF was supported by the \FNRS under the excellence of science (EOS) be.h project \no 30820817.
The work of JH was partially supported by Portuguese Fundação para a Ciência e a Tecnologia (FCT) through the projects CFTP-FCT Unit UIDB/\allowbreak00777/\allowbreak2020 and UIDP/\allowbreak00777/\allowbreak2020.
Computational resources have been provided by the supercomputing facilities of the Université catholique de Louvain (CISM/UCL) and the Consortium des Équipements de Calcul Intensif en Fédération Wallonie Bruxelles (CÉCI) funded by the \FNRS under convention 2.5020.11 and by the Walloon Region.

\appendix

\section{\Run 3 results} \label{sec:run 3 results}

This section presents the results for \Run 3 of the \LHC.
In \cref{fig:run 3 datasets}, we show the comparison between the trigger scenarios for each of the datasets; in \cref{fig:run 3 trigger}, we show the comparison between the datasets for each of the trigger scenarios.
The corresponding integrated luminosities of the different types of datasets are given in \cref{tab:datasets}.

\section{Codes and cards} \label{sec:code}

This section presents the code used to generate the events used in this analysis.

\paragraph{Signal}

The \ALP signal presented in \cref{fig:feynman diagram} is generated in the $s$-channel and decayed to two muons using \software{MadGraph}
\begin{lstlisting}
define p = g u c d s b u~ c~ d~ s~ b~
p p > ax > mu+ mu-
set mb 0
set ymb 0
\end{lstlisting}
In order to use the \fiveFS the mass of the bottom quark and its coupling are set to be zero.
The \software{ALPsEFT} \cite{Brivio:2017ije} model is adjusted by changing the \nolinkurl{param.card} values
\begin{lstlisting}
CGtil = CaPhi = 0.00001
\end{lstlisting}
We ensure that all time of flight information are stored in the \LHE files by adjusting the \nolinkurl{run.card} using
\begin{lstlisting}
time_of_flight = 0.0mm
\end{lstlisting}
The \PDF for the \pp datasets is included via
\begin{lstlisting}
nn23lo1:230000
\end{lstlisting}
while the \PDF of the \Pb\Pb\ dataset is included via
\begin{lstlisting}
lhapdf:901300
\end{lstlisting}

\paragraph{Background}

For the background simulation \software{Pythia} is generally initialised with
\begin{lstlisting}
HardQCD:gg2bbbar
HardQCD:qqbar2bbbar
PhaseSpace:p_THatMin = 20
\end{lstlisting}
For the \pp simulations we use the default conditions, \ie
\begin{lstlisting}
Beams:idA = 2212
Beams:idB = 2212
Beams:eCM = 13600.0 # Run 3
Beams:frameType = 1
\end{lstlisting}
and for the heavy ion background simulation we use
\begin{lstlisting}
Beams:idA = 1000822080
Beams:idB = 1000822080
Beams:eCM = 5500.0 # Run 3
HeavyIon:mode = 1
\end{lstlisting}
The pileup events are generated using the \software{Delphes} \nolinkurl{minbias} generation card which utilise the \code{SoftQCD} switch of \software{Pythia}

\paragraph{Detector}

For the scouting dataset, the muon tracking resolution is degraded in the \software{Delphes} \CMS card using
\begin{lstlisting}
set ResolutionFormula{
  (abs(eta) <= 0.9) * (pt > 0.1 && pt < 50) * (0.01) +
  (abs(eta) > 0.9 && abs(eta) <= 1.2) * (pt > 0.1 && pt < 50) *
    (0.01 + (abs(eta) - 0.9) * 0.01 * (2/3)) +
  (abs(eta) > 1.2 && abs(eta) <= 1.9) * (pt > 0.1 && pt < 50) * (0.03) +
  (abs(eta) <= 0.5) * (pt > 50) * sqrt(0.01^2 + pt^2 * 1.0e-4^2) +
  (abs(eta) > 0.5 && abs(eta) <= 1.5) * (pt > 50) *
    sqrt(0.015^2 + pt^2 * 1.5e-4^2) +
  (abs(eta) > 1.5 && abs(eta) <= 2.5) * (pt > 50) *
    sqrt(0.025^2 + pt^2 * 3.5e-4^2)
}
\end{lstlisting}
corresponding to the arguments made in \ccite{CMS:2019buh}.

\bibliographystyle{JHEP}
\bibliography{main}

\end{document}

%% file: definitions.tex
\usepackage{hep-math}
\usepackage[oldstyle,size=default]{hep-font}
\usepackage{hep-text}
\usepackage{hep-math-font}
\usepackage{hep-float}
\usepackage{hep-reference}
\usepackage{hep-acronym}

\acronym{ALP}{axion-like particle}
\acronym{HNL}{heavy neutral lepton}
\acronym{MC}{Monte Carlo}
\acronym{LHC}{Large Hadron Collider}
\acronym[$pp$]{pp}{proton-proton}
\acronym[HL-LHC]{HLLHC}{high luminosity phase of the Large Hadron Collider}
\shortacronym{LHCb}{Large Hadron Collider beauty}
\shortacronym{CMS}{Compact Muon Solenoid}
\shortacronym{ATLAS}{A Toroidal \LHC ApparatuS}
\shortacronym{CDF}{Collider Detector at Fermilab}
\acronym[5FS]{fiveFS}{five flavour scheme}
\acronym{PDF}{parton distribution function}
\acronym{IPS}{impact parameter significance}
\acronym{HLT}{high level trigger}
\acronym{LLP}{long-lived particle}
\acronym{NLO}{next-to-leading order}
\acronym{QCD}{quantum chromodynamics}
\acronym{SM}{Standard Model}
\acronym{EFT}{effective field theory}[effective field theories]
\acronym{BSM}{beyond the Standard Model}
\acronym{CL}{confidence level}
\acronym{EW}{electroweak}
\acronym{EWSB}{electroweak symmetry breaking}
\acronym{HI}{heavy ion}
\acronym{CM}{center of mass}
\acronym{PNGB}{(pseudo) Nambu-Goldstone boson}
\acronym{SMEFT}{Standard Model effective field theory}[Standard Model effective field theories]
\acronym[L1]{Lone}{level-1}
\acronym{LHE}{Les Houches event}
\longacronym{NP}{new physics}
\acronym[F.R.S.-FNRS]{FNRS}{Fond de la Recherche Scientifique de Belgique}

\usepackage[feynman,plot]{hep-graphic}
\graphicpath{./graphics}
\input{graphics/plots}
\pgfplotsset{table/search path={data}}



\let\U\relax\DeclareMathOperator{\U}{U}
\newcommand{\Run}[1]{Run~#1}
\newcommand{\pT}[1][]{\ensuremath{p_T^{#1}}\xspace}
\newcommand{\invfb}[1][]{\ensuremath{\unit[#1]{\inv{fb}}}\xspace}
\newcommand{\ion}[1]{\textrm{#1}}
\newcommand{\Pb}{\ion{Pb}}
\renewcommand{\O}{\ion{O}}

\usepackage{listings}

%% file: graphics/plots.tex
\usepackage{units}
\usepackage{amstext}

\newcommand\datasets{
  scouting/Scouting,
  bpark/Parking,
  PU/Standard $pp$,
  noPU/Low-pileup,
  HI/Heavy ion
}

\newcommand\triggers{
  single/serr/Muon,
  hybrid/herr/Hybrid,
  double/derr/Dimuon
}

\newcommand\nodata{
  bpark/double,
  scouting/single,
  scouting/hybrid,
  HI/single,
  HI/hybrid,
}

\newif\ifdraw

\pgfplotscreateplotcyclelist{trigger}{solid, dashed, {densely dotted, semithick}}
\pgfplotscreateplotcyclelist{datasets}{blue, red, green!50!black, violet, orange}

\pgfplotsset{ %
  data/.style={
    ylabel=$P_\text{norm}$,
    const plot mark mid,
    no marks,
    legend style={
      at={(0,1)},
      anchor=north west,
      outer sep=1ex,
      fill=none,
    },
  },
  signal/.style={
    data,
    colormap/jet,
    cycle list={[samples of colormap={9}]},
    legend columns=2,
    transpose legend,
    enlarge y limits={
      value=.25,
      upper,
    },
  },
  background/.style={
    data,
    cycle list name=datasets,
    ymax=1,
    legend style={
      at={(0,1)},
      anchor=north west,
    },
  },
  mass/.style={
    xlabel=$m_{\mu\mu}/\unit{GeV}$,
    ymin=1e-3,
  },
  d0/.style={
    xlabel=$d_0/\unit{mm}$,
    ymin=1e-3,
  },
}

\newcommand{\addsignalplot}[2]{%
  \addplot table[y expr=\thisrow{m#2}+1e-9]{#1};
  \expandafter\addlegendentry\expandafter{#2}
}

\newcommand{\addbackgroundplot}[3]{%
  \addplot table[y expr=\thisrow{#2}+1e-9]{#1};
  \expandafter\addlegendentry\expandafter{#3}
}

\pgfplotsset{ %
  results/.style={
    xlabel=$m_{\mu\mu}/\unit{GeV}$,
    ylabel={Significance in arbitrary units},
    legend style={
      at={(1,0)},
      anchor=south east,
      outer sep=1ex,
      fill opacity=.9,
      draw opacity=1,
      text opacity=1,
    },
    cycle multi list={datasets\nextlist trigger},
  },
  trigger/.style={
    three panels,
    results,
  },
  datasets/.style={
    three panels,
    results,
  },
  run/.style={
    results,
    legend columns=6,
    transpose legend,
  },
  error bars/.style={
    error bars/.cd,
    x dir=both,
    x explicit,
    y dir=both,
    y explicit,
    error bar style={solid}
  },
  run legend/.style={
    axis y line=none,
    axis x line=none,
    legend style={
      at={(.55,0)},
      inner sep = 0.5ex,
    },
  },
}

\newcommand{\addtriggerplot}[4]{
  \addplot+[error bars] table [y=#2, y error=#3] {#1.dat};
  \expandafter\addlegendentry\expandafter{#4}
}

\newcommand{\checkdraw}[2]{
  \global\drawtrue
  \foreach \nodataset/\notrigger in \nodata {\ifthenelse{\equal{#1}{\nodataset}\AND\equal{#2}{\notrigger}}{\global\drawfalse}{}}
}

\newcommand{\addtriggerplots}[2]{
  \addlegendtitle{\emph{Triggers}}
  \setcounter{cyclelistshift}{0}%
  \foreach \dataset/\datasetname in \datasets {
    \foreach \trigger/\error/\triggername in \triggers {
      \ifthenelse{\equal{#2}{\dataset}}{
       \checkdraw{\dataset}{\trigger}
        \ifdraw \addtriggerplot{\dataset_#1_new}{\trigger}{\error}{\triggername} \else \cyclelistshift \fi
      }{\cyclelistshift}
    }
  }
}

\newcommand{\adddatasetplot}[4]{
  \addplot+[error bars] table [y=#2, y error=#3] {#1.dat};
  \expandafter\addlegendentry\expandafter{#4}
}

\newcommand{\adddatasetplots}[2]{
  \addlegendtitle{\emph{Datasets}}
  \setcounter{cyclelistshift}{0}%
  \foreach \dataset/\datasetname in \datasets {
    \foreach \trigger/\error/\triggername in \triggers {
      \ifthenelse{\equal{#2}{\trigger}}{
       \checkdraw{\dataset}{\trigger}
        \ifdraw \adddatasetplot{\dataset_#1_new}{\trigger}{\error}{\datasetname} \else \cyclelistshift \fi
      }{\cyclelistshift}
    }
  }
}

\newcommand\addrunlegend{
\begin{axis}[
  run legend,
  cycle list name = trigger,
  legend style={anchor=south east},
]
  \setcounter{cyclelistshift}{0}%
  \addlegendtitle{}
  \addlegendtitle{}
  \addlegendtitle{\emph{Triggers}}
  \foreach\trigger/\error/\triggername in \triggers{
    \addplot coordinates {(0,0)};
    \expandafter\addlegendentry\expandafter{\triggername}
  }
\end{axis}
\begin{axis}[
  run legend,
  cycle list name = datasets,
  legend style={anchor=south west},
]
  \setcounter{cyclelistshift}{0}%
  \addlegendtitle{\emph{Datasets}}
  \foreach \dataset/\datasetname in \datasets{
    \addplot coordinates {(0,0)};
    \expandafter\addlegendentry\expandafter{\datasetname}
  }
\end{axis}
}